\documentclass[aip,apl,graphicx,reprint]{revtex4-1}

\usepackage{graphicx}
\usepackage{dcolumn}
\usepackage{bm}

\usepackage[utf8]{inputenc}
\usepackage[T1]{fontenc}
\usepackage{mathptmx}
\usepackage{amsmath,xcolor,colortbl,multirow,chngcntr,float,xr}
\usepackage{booktabs}
\usepackage{multirow}

\begin{document}

\preprint{AIP/123-QED}

\title[]{Electronic Structure and Stability of Cs$_2$TiX$_6$ and Cs$_2$ZrX$_6$ (X = Br, I) Vacancy Ordered Double Perovskites}
\author{B. Cucco}
\affiliation{Univ Rennes, ENSCR, INSA Rennes, CNRS, ISCR (Institut des Sciences Chimiques de Rennes), UMR 6226, France}
\author{G. Bouder}
\affiliation{Univ Rennes, ENSCR, INSA Rennes, CNRS, ISCR (Institut des Sciences Chimiques de Rennes), UMR 6226, France}
\author{L. Pedesseau}
\affiliation{Univ Rennes, INSA Rennes, CNRS, Institut FOTON - UMR 6082, Rennes, France}
\author{C. Katan}
\affiliation{Univ Rennes, ENSCR, INSA Rennes, CNRS, ISCR (Institut des Sciences Chimiques de Rennes), UMR 6226, France}
\author{J. Even}
\affiliation{Univ Rennes, INSA Rennes, CNRS, Institut FOTON - UMR 6082, Rennes, France}
\author{M. Kepenekian}
\affiliation{Univ Rennes, ENSCR, INSA Rennes, CNRS, ISCR (Institut des Sciences Chimiques de Rennes), UMR 6226, France}
\author{G. Volonakis}
\email[]{yorgos.volonakis@univ-rennes1.fr}
\affiliation{Univ Rennes, ENSCR, INSA Rennes, CNRS, ISCR (Institut des Sciences Chimiques de Rennes), UMR 6226, France}


\begin{abstract}
Vacancy ordered halide perovskites have been extensively investigated as promising lead-free alternatives to halide perovskites for various opto-electronic applications. Among these Cs$_{2}$TiBr$_{6}$ has been reported as a stable absorber with interesting electronic and optical properties, such as a band-gap in the visible, and long carrier diffusion lengths. Yet, a thorough theoretical analysis of the exhibited properties is still missing in order to further assess its application potential from a materials design point of view. In this letter, we perform a detailed analysis for the established Ti-based compounds and investigate the less-known materials based on Zr. We discuss in details their electronic properties and band symmetries, highlight the similarity between the materials in terms of properties, and reveal limits for tuning electronic and optical properties within this family of vacancy ordered double perovskites that share the same electron configuration. We also show the challenges to compute accurate and meaningful quasi-particle corrections at GW level. Furthermore, we address their chemical stability against different decomposition reaction pathways, identifying stable regions for the formation of all materials, while probing their mechanical stability employing phonon calculations. We predict that Cs$_{2}$ZrI$_{6}$, a material practically unexplored to-date, shall exhibit a quasi-direct electronic band-gap well within the visible range, the smallest charge carrier effective masses within the Cs$_{2}$BX$_{6}$ (B=Ti,Zr; X=Br, I) compounds, and a good chemical stability.
\end{abstract}

\maketitle

Over the past decade lead-based halide perovskites have revolutionized the field of emerging photovoltaic technologies, reaching record-breaking power conversion efficiencies (PCE) certified at above 25\%~\cite{Jeong2021,nrel}. To-date, these perovskites have also found a wide range of opto-electronic applications in various devices such as highly-efficient light-emitters~\cite{Cui2020}, photo-catalysts~\cite{Zhang2019}, and X-ray detectors~\cite{Li2019}. In terms of materials, there have been important efforts to increase the exhibited stability while at the same time replacing the non-environmentally friendly lead atoms, which led to the exploration of different candidate perovskite materials~\cite{Giustino2016}. For example, within the APbX$_3$ corner-sharing octahedra lattice it is possible to retain the oxidation state at the metallic site by replacing Pb$^{\textrm{2+}}$ with Sn$^{\textrm{2+}}$ \cite{Shao2018,Lintao2020}. Sn$^{\textrm{2+}}$ based perovskite materials exhibit similar electronic and optical properties than lead halide perovskites, yet, ASnX$_3$ perovskites have a poor stability and degrade fast~\cite{Noel2014,Hao_2014} due to the rapid oxidation of Sn$^{\textrm{2+}}$ into Sn$^{\textrm{4+}}$. Another strategy is to replace Pb$^{\textrm{2+}}$ using atoms at different oxidation states, and hence increase significantly the materials search space~\cite{Giustino2016}. The combination of monovalent and trivalent atoms at the B-site lead to the formation of the A$_2$BB$^{\prime}$X$_6$ halide double perovskites, such as Cs$_2$BiAgBr$_6$~\cite{Volonakis2016,Slavney2016} and Cs$_2$InAgCl$_6$~\cite{Volonakis2017}. Using trivalent atoms lead to the formation of A$_3$B$_2$X$_9$ structures like Cs$_3$Bi$_2$I$_9$~\cite{Chabot1978,Bai2018}, while employing tetravalent atoms can lead to the so-called  vacancy ordered double perovskite (VODP) A$_2$BX$_6$ lattices like Cs$_2$SnI$_6$. In fact, oxidation of Sn$^{2+}$ in CsSnI$_3$ due to exposure to air, can lead to a structural transformation to Cs$_2$SnI$_6$~\cite{Xiaofeng2017} with Sn$^{4+}$. 

The crystal lattice of such VODP can be derived from the standard A$_2$BB$^{\prime}$X$_6$ double perovskite lattice by periodically introducing vacancies at the B$^{\prime}$ site. The final A$_2$BX$_6$ structure consists on a checkerboard like pattern between vacancies and isolated BX$_6$ octahedra. VODP materials can be synthesized with various tetravalent metals at the B-site including Pd~\cite{Sakai2017}, Te~\cite{Maughan2016}, Pt~\cite{Evans2018}, Ti~\cite{Chen2018} and the aforementioned group 14 atoms: Pb~\cite{Engel1933} and Sn~\cite{Kaltzoglou2016}.
Many of these materials are non-toxic, stable and with tunable electronic and optical properties, making them promising candidates for various optoelectronic applications~\cite{Ju2018,Maughan2019}. Cs$_2$SnI$_6$ for example is found to be more stable than corner-sharing CsSnI$_3$ and is the first A$_2$BX$_6$ material employed in solar cell devices~\cite{Xiaofeng2017}. Ti-based compounds Cs$_2$TiX$_6$ (X=Br, I) have been successfully synthesized and exhibit band-gaps of 1.02-2.00 eV~\cite{Ju2018,Chen2018,Euvrard2020}. Photovoltaic devices using Cs$_2$TiBr$_6$ have achieved PCE of up to $3.3\%$ in a planar-heterojunction architecture with C$_{60}$, exhibiting open-circuit voltages (V$_{\textrm{oc}}$) of 1.0~V and also a short-circuit current (J$_{sc}$) of 5.69 mA.cm$^{\textrm{-2}}$ ~\cite{Chen2018}. The relatively poor PCE of the photovoltaic devices has been attributed to low absorption, indirect nature of the band-gap and the intrinsic instability of the material under ambient conditions~\cite{Euvrard2020}. 

\begin{figure*}[!ht]
 \begin{center}
  \includegraphics[width=1\textwidth]{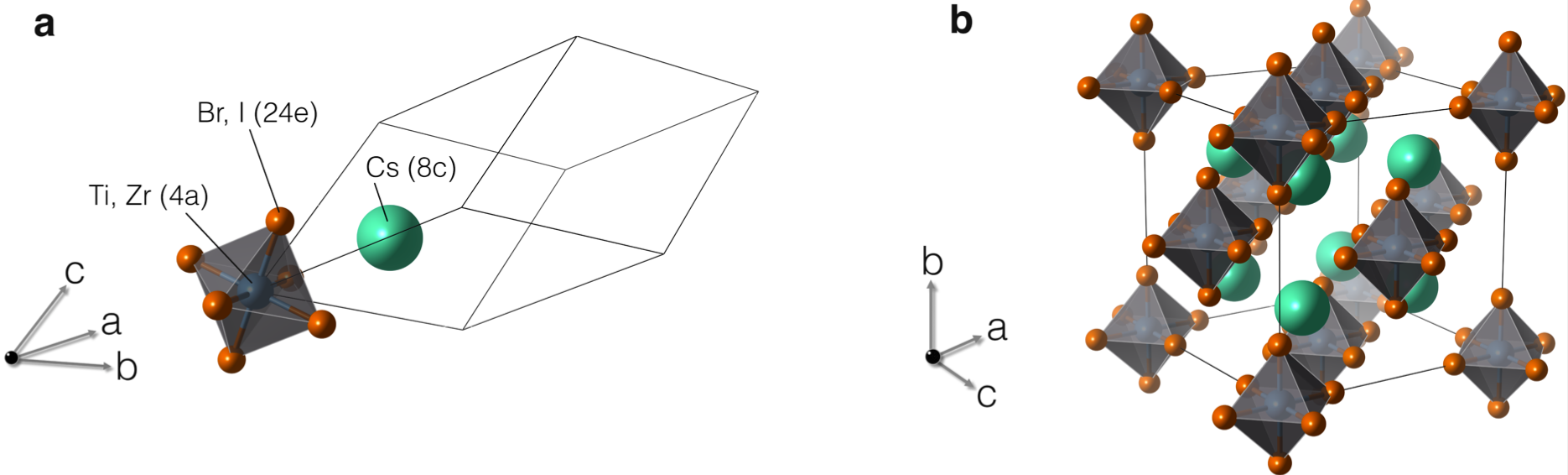}
 \end{center}
 \caption{(a) Primitive unit cell of the VODP family with corresponding Wyckoff positions. (b) Conventional unit cell showing the rock-salt arrangement between BX$_{6}$ octahedra and vacancies}
 \label{fig:1}
\end{figure*}

In this letter, we perform a thorough analysis of the electronic structure, mechanical and chemical stability, and optical properties, of the class of photo-active Cs$_2$BX$_6$ VODP materials with B-site cations Ti$^{+4}$ and Zr$^{+4}$, hence with d$^{\textrm{0}}$ valency at the B-site. Doing so, we explore the potential tunability limits of this family of materials, extend the existing understanding of the well-established Cs$_2$TiX$_6$; X=Br, I, and investigate the less known Zr-based compounds Cs$_2$ZrX$_6$, X=Br, I. We address the electronic and optical properties of these materials by means of density functional theory calculations, where we perform a complete symmetry analysis of the band structures and reveal its consequences on important physical properties like the exhibited band-gaps and charge carrier effective masses. Furthermore, we follow different decomposition reaction pathways and show the relative stability of the VODP materials with respect to competing compounds, and finally we address their mechanical stability by means of phonon dispersion calculations. Throughout this work we will discuss several aspects regarding both Ti and Zr based materials, and as the physical and chemical properties are very similar between these, in the main text we include the results for the known stable Cs$_{2}$TiBr$_{6}$ and the newly proposed Cs$_{2}$ZrI$_{6}$, while the results for the other materials are shown in the SI.

Vacancy ordered double perovskites crystallize in a $Fm\bar{3}m$ face-centered cubic lattice (space group: 225), that can be considered as a rock-salt arrangement of BX$_6$ and $\Delta$X$_6$ octahedra, with $\Delta$ a vacancy site. The primitive and conventional unit cells are shown in Fig.~\ref{fig:1}a and b, respectively. To begin, we fully optimize the four compounds Cs$_2$BX$_6$ (B=Ti,Zr; X=Br, I) within DFT-PBE (computational details are in the Supplementary Material - SI). Similar to standard perovskites, bromides have smaller lattice constants compared to the respective iodides. The calculated lattice parameters for Cs$_2$TiBr$_6$ and Cs$_2$ZrBr$_6$ are 10.92~\AA\ and 11.24~\AA, respectively. 
For the iodides Cs$_2$TiI$_6$ and Cs$_2$ZrI$_6$ we obtain larger lattice constants of 11.75~\AA\ and 12.00~\AA. These are in good agreement with experimental values of 10.57-10.92~\AA\ for Cs$_2$TiBr$_6$~\cite{Kong2020,Euvrard2020,Ju2018}, and 11.67~\AA\ for Cs$_2$TiI$_6$~\cite{Ju2018}. The B-X bond-lengths follow the same trend, but interestingly the size of the vacancy doesn't seem to be significantly affected by the ionic radii of the B-site atom. We can quantify the size of the vacancy site as half the distance between two halogens that bound to a vacancy (i.e., X-$\Delta$-X), and find 2.96~\AA, 2.97~\AA, 3.10~\AA, 3.12~\AA, for Cs$_2$TiBr$_6$, Cs$_2$ZrBr$_6$, Cs$_2$TiI$_6$, and Cs$_2$ZrI$_6$ respectively. This is in fact consistent with the fact that only halogens surround the vacant site, hence their interactions define the size of this $\Delta$X$_6$ octahedron. 


\begin{figure*}[!ht]
 \begin{center}
  \includegraphics[width=0.9\textwidth]{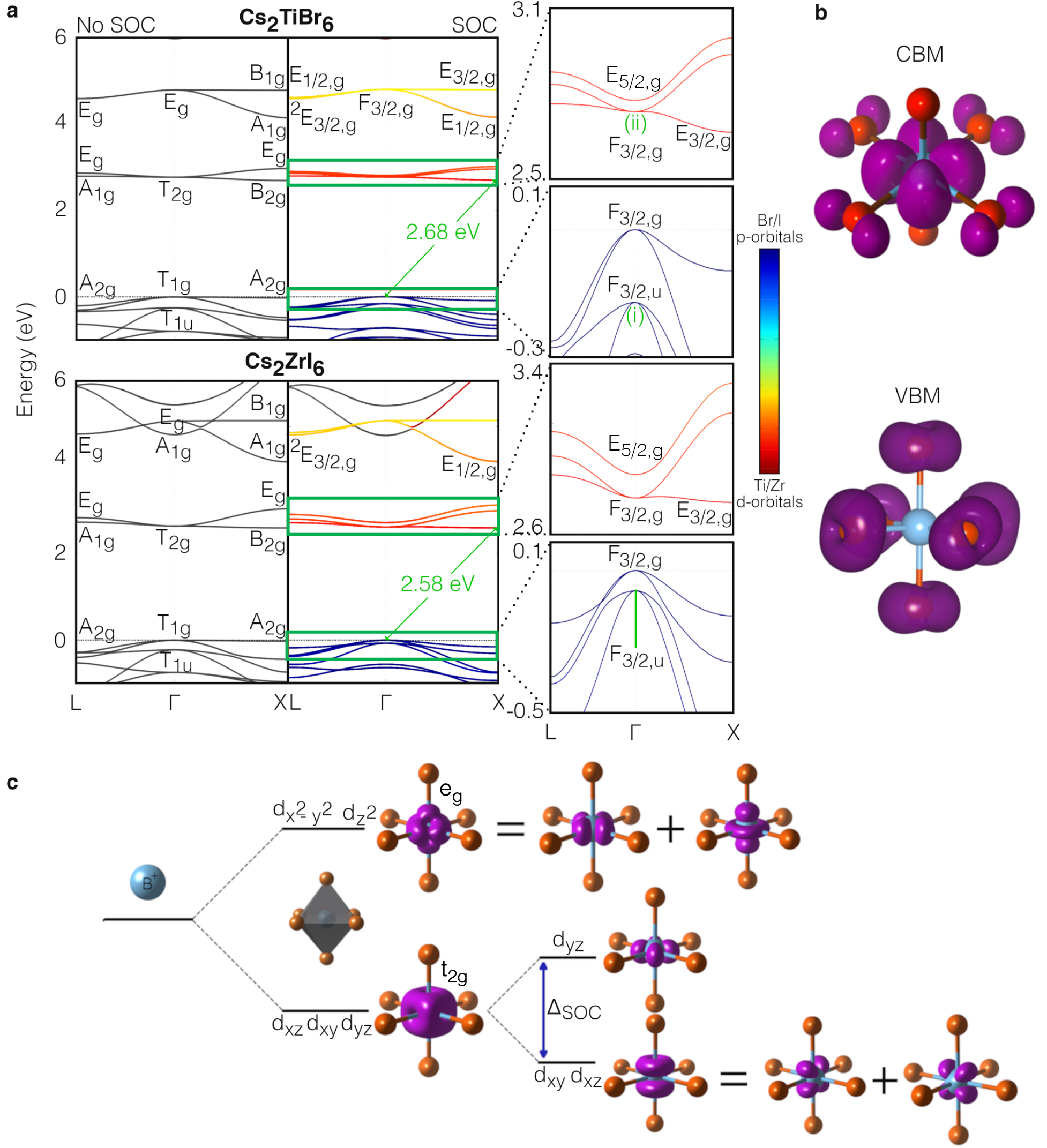}
 \end{center}
 \vspace{-1.0cm}
 \caption{(a) Electronic band structure and symmetry analysis with and without spin-orbit coupling (SOC) for Cs$_{2}$TiBr$_{6}$ and Cs$_{2}$ZrI$_{6}$. Valence band maximum is set to zero, conduction bands are rigidly shifted to match the DFT-HSE level. Colors on the SOC electronic band structures denote the projection of the halogen p-orbitals (blue) and B-site atoms d-orbitals (red) to the respective state. The complete character table and the notation for the irreducible representations are in Table S3. (b) Pseudo-charge density for the  valence band maximum (VBM) and conduction band minimum (CBM). (c) Splitting of Ti/Zr d-states due to octahedral coordination and $\Delta_\textrm{{SOC}}$ split.}
 \label{fig:2}
\end{figure*}

Next, we look at the electronic properties and symmetry of the VODPs. To do so, we employ and discuss below three levels of theory: standard DFT-PBE, HSE06 hybrid functional~\cite{Heyd2003} (DFT-HSE), and finally G$_\textrm{o}$W$_\textrm{o}$ (GW) calculations. Figure~\ref{fig:2}a shows the electronic band structures of Cs$_2$TiBr$_6$ and Cs$_2$ZrI$_6$, with and without spin-orbit coupling (SOC) effects. We also include the irreducible representations close to the band edges. The band structures of the two compounds share similar features, with the Cs$_2$ZrI$_6$ exhibiting slightly larger bandwidths. Not surprisingly, the electronic structure of the VODP is dictated by the valency of the B$^{4+}$-site and its interaction with X-site $p$-orbitals, which is in fact similar to the case of single~\cite{Mao2018} and double perovskites~\cite{ Volonakis2017b, Volonakis2019,Leveillee2021}. Here, the conduction and valence band edges are comprised of $d$-orbitals from the B-site atoms and $p$-orbitals from the halogens (projection of the orbitals on the bands are shown in Fig.~\ref{fig:2}a and Fig.~S1). The electronic band-gaps are indirect with the conduction band minimum (cbm) at the X-point of the FCC Brillouin zone, and the valence band maxima (vbm) at the $\Gamma$ point. The pseudo-charge density plots are shown in Fig.~\ref{fig:2}b. Within DFT-HSE, we calculate electronic band-gaps of 2.68~eV and 2.58~eV for Cs$_2$TiBr$_6$ and Cs$_2$ZrI$_6$, respectively (see Table S1 for all compounds). Smaller halogens exhibit larger band-gaps, and the smallest the B-site atom the smallest the band-gap. This is consistent with previous reports for A$_2$BX$_6$ (A=Cs,K,Rb; B=Sn,Pt,Te; X=Cl,Br,I)~\cite{Cai2017} and Cs$_2$SbX$_6$ (X=Cl,Br)\cite{Day1963}. Comparing Cs$_{2}$TiBr$_{6}$ and Cs$_{2}$ZrI$_{6}$, within DFT-PBE and DFT-HSE, we note an inversion of the band-gap trend (i.e., the DFT-HSE band-gap of Cs$_{2}$ZrI$_{6}$ becomes smaller than the band-gap of Cs$_{2}$TiBr$_{6}$), which is due to the larger exchange correction for the bromide materials with respect to the one for iodides in the hybrid xc-functional calculations. Overall, comparing the accuracy of the DFT-HSE calculations with the experimental reports one can observe a significant band-gap overestimation for Cs$_2$TiBr$_6$ and Cs$_2$TiI$_6$. This can relate to either the direct comparison between measured optical band-gaps with the calculated electronic band-gaps, or to the presence of defects in the samples used for the measurements. On the other hand, for the case of Cs$_{2}$ZrBr$_{6}$ DFT-HSE predicts an electronic band-gap just 0.12~eV above the measured optical gap~\cite{abfalterer2020}. To address these effects, further thorough experimental work needs to be carried alongside an ab-initio study including electron-hole interactions. 

\begin{figure*}[!ht]
 \begin{center}
  \includegraphics[width=0.65\textwidth]{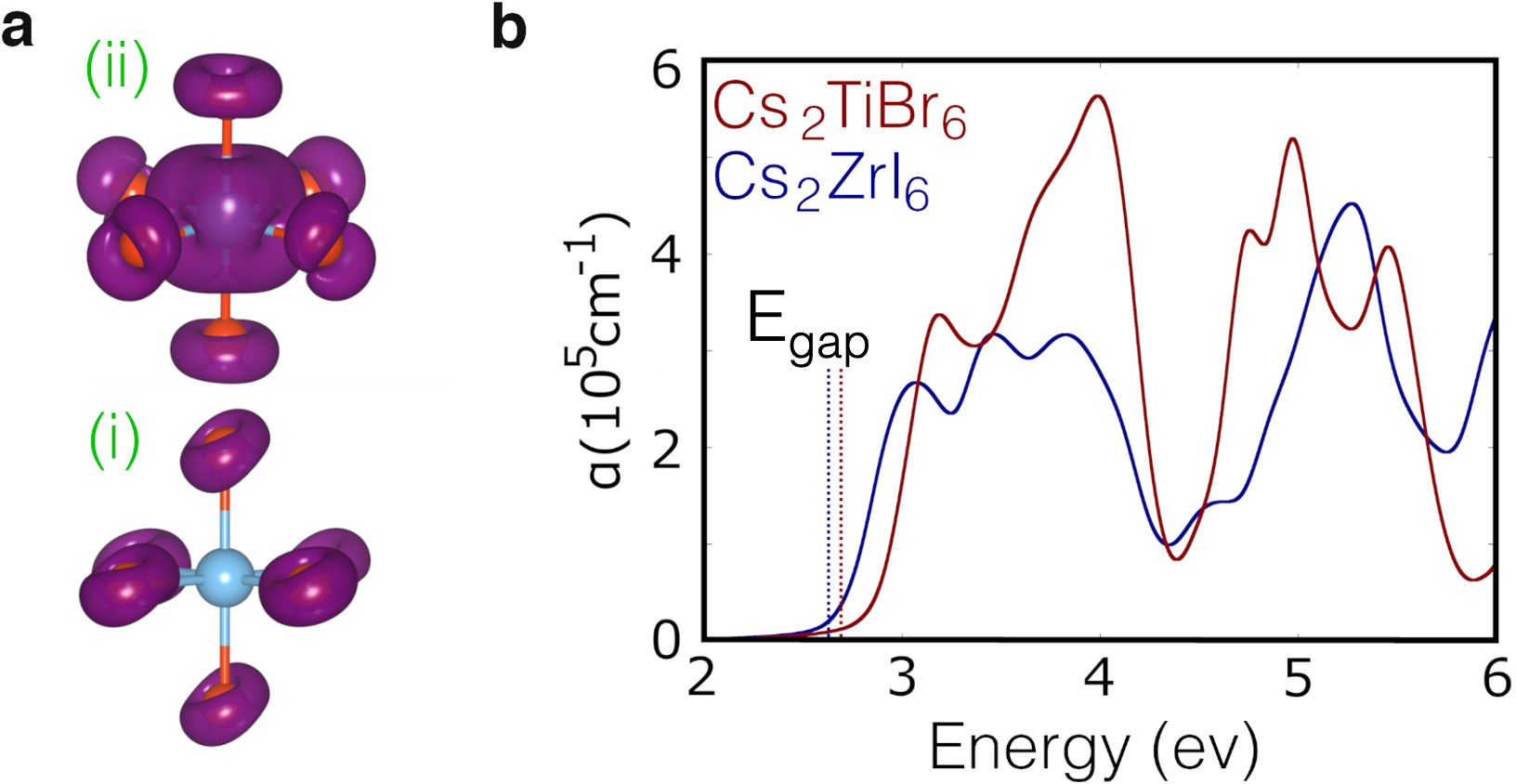}
 \end{center}
 \caption{(a) Pseudo-charge density for the first symmetry allowed transition between points (i) to (ii) as shown in Fig.~\ref{fig:2}a. (b) Calculated absorption coefficient with the DFT-HSE electronic band-gap at the dotted lines. }
 \label{fig:3}
\end{figure*}

The calculated GW quasi-particle gaps are summarized in Table~S1. For Cs$_{2}$TiBr$_{6}$ the predicted 3.87~eV quasi-particle gap overestimates more than two times the experimentally measured optical band-gap of 1.78-2.00~eV ~\cite{Chen2018,Ju2018,Euvrard2020}, and is significantly larger than our HSE calculation value of 2.68~eV.  In fact, GW significantly overestimates the band-gaps of all three materials, for which experimental data are available (see Table~S1). Such overestimation, can be attributed to the DFT starting point of our GW calculation and the presence of localized $d$-orbitals at the band edges with strong correlation effects, which standard DFT xc-functionals fail to describe. This effect can be clearly shown by comparing our GW and HSE interpolated band structures (Fig.~S2), where it is clear that there are practically no differences for the description of the halogens $p$-orbitals and the $Cs$ $s$-orbitals. However, the same is not true for the relative position of Ti/Zr ${d}$-states indicating that indeed a careful treatment needs to be done in order to compute accurate quasi-particle band-gap for these materials. In fact, this is a well-known limitation, and to overcome it one can for example employ the auto-consistent DFT+U-GW method proposed by Patrick and Giustino~\cite{Patrick_2012}. Its application on TiO$_2$, another system for which Ti $d$-orbitals are critical for the electronic structure, corrected the DFT-GW overestimated electronic band-gap of 3.7~eV, to 3.27~eV. In our case, since the dispersion of the electronic band structures within PBE, HSE and GW does not change significantly (Fig.~S2 and S3) we employ the PBE xc-functional, shifting the conduction band states rigidly to match the DFT-HSE band-gap.

The cbm of the VODP is made of $d_{xy}$ of the Ti or Zr atoms, that interact with the p-orbitals of the halogens. As diagrammatically shown in Fig.~\ref{fig:2}c the split of the $d$-orbitals due to an octahedral coordination lead to an $e_g$ double degenerated $d_{x^{2}-y^{2}}$ and $d_{z^{2}}$ state and a $t_{2g}$ triple degenerated orbitals $d_{xy}, d_{xz}, d_{yz}$ state, exactly as we find at the $\Gamma$ point in the VODP band structures without SOC. On the same figure the charge density plots of these points are shown to be  consistent with a simple superposition of the aforementioned states, in agreement with the pseudo-charge density plots of Cs$_2$TiBr$_6$ and Cs$_2$ZrI$_6$. Inclusion of the SOC interactions is important for the conduction band of the VODP compounds and leads to a second splitting of $t_{2g}$ states at the cbm at $\Gamma$. We can further look at this splitting; for Cs$_2$TiBr$_6$ and Cs$_2$ZrI$_6$ the splits are 22~meV and 85~meV respectively, within DFT-HSE. $\Delta_{SOC}$ increases for compounds with heavier atoms on which the SOC effects are stronger. In fact,  the magnitude of the $\Delta_{SOC}$ split is matching the reference atomic $\Delta_{SOC}$ values of 47~meV and 155~meV, for Ti and Zr $d$-orbitals, respectively~\cite{NIST_ASD}. On the other hand, the vbm of all VODP materials is comprised exclusively from $p$-orbitals from the halogens, which arrange around the vacant site as shown in Fig.~S4.

The lowest direct band-to-band transition is in fact forbidden by symmetry (both at $\Gamma$ and at X-point). However, few meV below the vbm, 161~meV (67~meV)  Cs$_2$TiBr$_6$ (Cs$_2$ZrI$_6$), we find the irreducible representation F$_{3/2,u}$, which comprises the first symmetry allowed transition to the F$_{3/2,g}$ representation at the conduction band. The charge density corresponding for these states are shown in Fig.~\ref{fig:3}a. We note here, that inclusion of SOC is also important for the fine description of the optically active transitions of these VODP, since the irreducible representation related to the first symmetry allowed band-to-band transition without taking into account SOC is at 250~meV and 220~meV below the vbm for Cs$_2$TiBr$_6$ and Cs$_2$ZrI$_6$, respectively.  Figure \ref{fig:3}b shows the calculated absorption coefficient for these materials at RPA level. Here the absorption was obtained using DFT-PBE including SOC with a dense k-point grid of 16$\times$16$\times$16 and applying a scissor operator to match the HSE calculated gap. To verify this approach, in Fig.~S3 we compare the electronic structure of the scissor shifted PBE calculation with the HSE hybrid functional bands. It is clear that the features of the bands are well described so, one can expect that optical properties (i.e., only considering inter-band transitions) calculated from a scissor shifted DFT-PBE are in close agreement with hybrid functional optics. One can observe a small shift at the onset of the absorption coefficient spectra with respect to the position of the DFT-HSE band-gap (dotted lines in Fig.~\ref{fig:3}b), which is consistent with the symmetry forbidden direct transition at $\Gamma$. 

\begin{table*}[!ht]
\caption{Charge carrier effective masses in units of $m_{e}$. For the holes ($m_h$) at the valence band top, the masses are uniform and we include the heavy and light bands. For electrons ($m_e$) at the conduction band bottom there are two components in the effective mass tensor, a fast and a slow direction. (All masses are in $m_{e}$)}
\begin{ruledtabular}
\begin{tabular}{ccccc}
 Effective masses~($m_{e}$)&$m_{h}^{light}$&$m_{h}^{heavy}$&$m_{e}^{fast}$&$m_{e}^{slow}$\\ \hline
 Cs$_{2}$TiBr$_{6}$&$1.14$&$2.84$&$1.76$&$7.62$  \\
 Cs$_{2}$TiI$_{6}$&$0.62$&$1.55$&$0.98$&$8.91$\\
 Cs$_{2}$ZrBr$_{6}$&$1.12$&$2.82$&$1.27$&$11.29$\\
 Cs$_{2}$ZrI$_{6}$&$0.60$&$1.45$&$0.86$&$11.55$ \\
\end{tabular}
\end{ruledtabular}
\label{tab:1}
\end{table*}

To probe the potential transport properties of these compounds we calculate the charge carrier effective masses at the band edges, which are summarized on the Table~\ref{tab:1}. Looking at the valence band, all materials exhibit heavy and light holes at $\Gamma$. Clearly, iodine-based materials exhibit lower effective masses overall, while hole masses are practically the same between bromides and iodides. This is consistent with the projection of the vbm to solely halogen orbitals, shown in Fig.~S1. At the conduction band, Zr-based compounds exhibit lower electron effective masses, hence, halogens affect both electron and hole masses, and B-site atoms affect only the electron mass. The conduction edge at the $X$ high-symmetry point exhibits two in-plane light effective masses (fast direction) and a heavy mass (slow direction) orthogonal to these as shown in the effective mass tensor of Fig.~S5. The directions of the masses are highlighted with colored arrows in Fig.~S6, with the fast component expanding through the charge density lobes (blue lines). The in-plane expansion of the charge density, is in fact 3-fold degenerate along the symmetry equivalent planes due to the cubic lattice of our systems, yet it could allow confining the fast carriers to a two-dimensional plane by introducing small octahedral distortions that would break the crystal symmetry~\cite{Salluzzo2009}.

Having established the electronic structure of this family of VODP, and since we identify Cs$_{2}$ZrI$_{6}$ as a potential good candidate material, we move on to explore its stability. In fact we also highlight two reports of the early eighties, where  Cs$_{2}$ZrI$_{6}$ has been reportedly synthesized~\cite{Guthrie_1982,sinram1982hexa}, though has since remained  unexplored. In an effort to understand the conditions for the formation of the material and compare with the rest of the VODP family, we first investigate their stability with respect to different decomposition reaction pathways. Following the procedure previously described by Persson \textit{et al.}~\cite{Persson2005}, firstly we imposed that the atomic chemical potentials $\mu_{i}$ of atom $i$ must be less or equal to its solid phase $\mu_{i}^{solid}$ in order to assure that there will be no precipitation of these, that is:
\begin{equation}
    \mu_{i} \le \mu_{i}^{solid} \Longrightarrow \Delta\mu_{i} \le 0,
\end{equation}
where $\mu_{i}=\mu_{i}^{solid}$ indicates the atom $i$-rich limit of the phase diagram. Secondly, the stability condition of the VODP Cs$_{2}$BX$_{6}$ is taken into account through:
\begin{equation}
\label{eq1}
2\Delta\mu_{Cs} + \Delta\mu_{B} + 6\Delta\mu_{X} = \Delta H(Cs_{2}BX_{6}), 
\end{equation}
where B=Ti, Zr, X=Br, I and $\Delta H \textrm{(Cs}_{2}\textrm{BX}_{6})$ is the heat of formation of compound Cs$_{2}$BX$_{6}$. Thirdly, we derive a set of equations which address the stability of all competing phases. Imposing that the sum of the chemical potentials of constituent atoms is less than the heat of formation of these means that there will be no formation of these compounds. Therefore, the analysis of the common solution of these set of inequalities lends to the thermodynamically stable region on which the compound of interest forms with no competition. The structures of all the stable competing compounds were identified using the Materials Project database~\cite{Jain2013} and for each material of the VODP family the set of inequalities (complete set in the SI) have the form:
\begin{equation}
m\Delta\mu_{Cs} + n\Delta\mu_{B} + l\Delta\mu_{X} \le \Delta H(Cs_{m}B_{n}X_{l}).
\end{equation}

\begin{figure*}[ht]
 \begin{center}
  \includegraphics[width=0.9\textwidth]{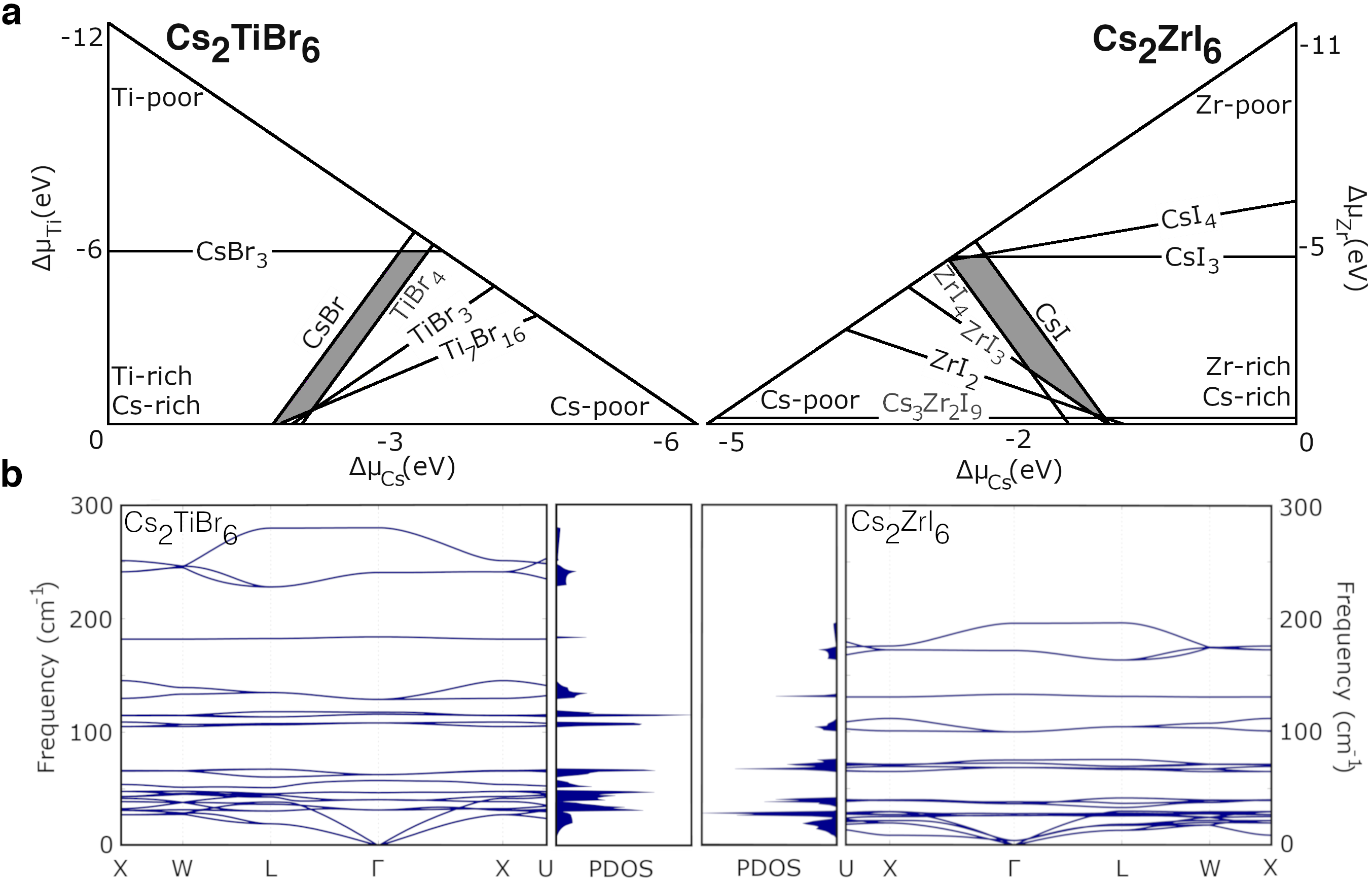}
 \end{center}
 \caption{(a) Decomposition pathway diagrams indicating all the known stable competing phases with stability region represented as the grey shaded area. (b) Phonon dispersion and phonon density of states, the absence of imaginary(or soft) modes indicates dynamical stability}
 \label{fig:4}
\end{figure*}

The obtained stable chemical potential limits for each of the four explored materials  are shown in Fig.~\ref{fig:4}a and Fig.~S7a as the grey shaded area. First, we find that all binary compounds that do not contain $Cs$ are stable within the $Cs-poor$ region of the diagrams while the ones containing $Cs$ are placed within the $B-poor$ limit, which shows the consistency of these diagrams.  Regarding the stability with respect to VODP decomposition to their competing phases, all plots show a region for which the materials are predicted to be stable (grey shaded). The position of the region is practically in the middle of the plots; hence we expect these to be more easily synthesized on a balanced chemical environment. A comparison between the size of these regions reveals that the Zr family of compounds are expected to be more stable against decomposition when compared to the Ti family, and within each of these families the materials containing iodine are expected to be less stable with respect to bromides. We also address the mechanical stability of the explored structural phases by computing the phonon dispersion relations within DFT-PBE. The absence of modes with imaginary frequencies confirms that these materials are indeed stable, as shown on Fig.~\ref{fig:4}b and Fig.~S7b. In fact, just recently A. Abfalterer \textit{et al.} successfully synthesized the Cs$_{2}$ZrCl$_{6}$ and Cs$_{2}$ZrBr$_{6}$ which was shown to be indeed stable, and with the bromide material exhibiting a measured absorption onset at 3.76~eV matching well the DFT-HSE predicted band-gap~\cite{abfalterer2020}.

To conclude, within this work we investigated the physico-chemical properties of Zr- and Ti-based VODP materials. We found that the not corner-sharing $BX_{6}$ octahedra effectively behave fundamentally like isolated structures, exhibiting the very same type of band splitting of tetrahedrally coordinated Zr and Ti atomic species as predicted within the crystal field theory. Yet, the charge carrier effective masses reported can be relatively low (i.e., below 1m$_e$), especially for the case of iodides. The importance of SOC interactions for the description of the allowed optical transitions is shown, together with a detailed symmetry analysis of the electronic band structure for the most prominent compounds in this VODP family: well-known Cs$_2$TiBr$_6$ and newly proposed Cs$_2$ZrI$_6$. Furthermore, we probe the limitation of ab-initio calculations and show that in order to compute meaningful quasi-particle corrections one needs to careful treat the conduction edge states due to strong correlations effects of $d$-orbitals, which can lead to largely overestimated band-gaps at the $GW$ level. The optical absorption is very similar between these materials, due to their similar electronic band structures, hence alloying these could be an interesting parameter to fine tune the optical onset, without changing the rest exhibited properties. Finally, all compounds are predicted to be stable with respect to decomposition to competing phases and show that all VODP within this family are mechanically stable. Overall, this work explains the atomistic details of the electronic and optical properties of the VODP family with $d^{0}$ valency, within which Cs$_{2}$ZrI$_{6}$ remains unexplored to-date, and  exhibits a band-gap within the visible range and the lightest charge carrier effective masses. Our findings support the possibility of forming a stable zirconium-iodine VODP with photo-active properties. 

\vspace{-0.2cm}

\begin{acknowledgments}

\vspace{-0.5cm}

The research leading to these results has received funding from the Chaire de Recherche Rennes Metropole project, and from the European Union’s Horizon 2020 program, through a FET Open research and innovation action under the grant agreement No 862656 (DROP-IT). This work was granted access to the HPC resources of TGCC under the allocations 2020-A0100911434 and 2020-A0090907682 made by GENCI. We acknowledge PRACE for awarding us access to the ARCHER2, United Kingdom.
\end{acknowledgments}

\vspace{-0.2cm}

\clearpage
\bibliography{bibliography.bib}

\end{document}